# FROM MEMS DEVICES TO SMART INTEGRATED SYSTEMS

*O. Soeraasen\* and J. E. Ramstad\**

\*Department of Informatics, University of Oslo, P O Box 1080 Blindern, N-0316 Oslo
oddvar@ifi.uio.no, janera@student.matnat.uio.no

## ABSTRACT

The smart integrated systems of tomorrow would demand a combination of micromechanical components and traditional electronics. On-chip solutions will be the ultimate goal. One way of making such systems is to implement the mechanical parts in an ordinary CMOS process. This procedure has been used to design an oscillator consisting of a resonating cantilever beam and a CMOS Pierce feedback amplifier. The resonating frequency is changed if the beam is bent by external forces. The paper describes central features of this procedure and highlights the design considerations for the CMOS-MEMS oscillator. The circuit is used as an example of a "VLSI designer" way of making future integrated micromechanical and microelectronic systems on-chip. The possibility for expansion to larger systems is reviewed.

## 1. INTRODUCTION

The smart integrated systems of tomorrow will need to have a great number of MEMS devices in combination with extensive processing power from surrounding microelectronic circuits. The MEMS devices extend the systems with "eyes, ears and fingers" for contacting the environment. The microelectronic circuits in their turn enhance the functionality of the MEMS components by connecting the signals from the mechanical sensors to amplifiers and digital preprocessing. Effective means of combining the technologies are required.

Today, each MEMS processing facility typically has its own secrets and special features which limit interoperability. CMOS processes, that dominate the IC industry, are much more standardized, which has made both the foundry concept and second sourcing essential for making VLSI systems in cost-effective ways. This division in a design activity separated from the subsequent semiconductor processing has expanded the design community and accelerated the impact of IC systems. The designers of the current generation of MST systems have their background in physics, material technology or chemistry and are working very closely to the MEMS processing labs. The designers of the future are expected to have their background in computer science and ASIC design and will typically attack the system development task in another way than traditionally done, working on a conceptual level separated from processing details. A great part of such a design process consists of handling geometries. The ultimate goal for the designer would be to integrate both micromechanical and microelectronic elements on-chip to make real SoCs (System-on-Chip). Cheaper and more standardized implementation procedures and processes are needed.

One way of making combined systems is to implement the mechanical parts in an ordinary CMOS process together with traditional electronics. As an example of such an approach, a CMOS-MEMS oscillator has been designed. The feasibility and flexibility of the method will be highlighted in the following, together with pointing out certain design constraints and problems of the approach. In chapter 2 an overview of on-chip integration methods is given with emphasis on the specific CMOS-MEMS method. Chapter 3 describes the actual system which consists of a resonating cantilever beam in a Pierce oscillator configuration. Some experiences from this task are generalized and the implication and possibilities for designing future intelligent, heterogeneous systems are considered in chapter 4. Chapter 5 concludes the work.

## 2. ON-CHIP INTEGRATION OF MEMS AND CMOS

The great advantage of multi chip packaging of MEMS and CMOS is the ability to combine quite diverse processes and materials such as glass, plastic, Silicon or organic compounds. Despite the flexibility of that approach, the methods are costly and normally introduce heavy load impedances and large stray capacitances in comparison to on-chip solutions. Monolithic integration, on the other hand, can give easier handling and lower production costs, higher reliability, and reduced parasitics. Today on-chip integration is done in various ways where the MEMS is typically implemented either as pre-CMOS, intermediate-CMOS, or post-CMOS [1]. Each approach has its own advantages and drawbacks. With a post-CMOS procedure, the MEMS processing has to be done on the finished CMOS wafers. High temperature steps should be avoided not to destroy the metal layers. Low temperature deposition might require specific structural materials to be used, such as SiGe [2]. Generally, whole wafers have to be post processed, constraining the wafers to be of equal dimensions in the CMOS foundry as for the MEMS processing - which is seldom the case. Post processing single CMOS chips is not practical if a separate MEMS masking is needed.

Typical for all the common integration procedures is their limitation in versatility and dissemination. The different processing "modules" and variants would need large investments for each new upgrade which in most cases do not pay off. On the other side, it is striking to observe the enormous investments in the IC industry for developing standardized and powerful CMOS processes with still finer line dimensions and higher speed. To be able to take advantage of that development is an appealing thought.

By our work, we investigate how an ordinary CMOS process can be used for making both mechanical components, CMOS-MEMS, and the interconnected electronics. The approach which we have chosen is the newly established European version of the ASIMPS procedure [3] offered by CMP (Circuits Multi-Projets). The 0.25 μm ST7RF BiCMOS process from ST Microelectronics [4] is used. Circuits from the CMOS run are then post processed at Carnegie Mellon University (CMU), Pittsburgh [5], where the micromechanical parts of the design are released in a mask-less etch and release process. The etching can be done chip-wise on individual diced chips with no extra masks needed, meaning that the CMOS process can be run as a MPW (Multi Project Wafer).

A multi layer stack of metals and dielectrics is used to create the MEMS structures, as shown in Figure 1. This material selection limits the possibility of achieving a very high Youngs modulus compared to what can be obtained by using polysilicon or polydiamond as in pure MEMS processes. A typical value of Youngs modulus is 63GPa [6]. However, using laminated structures with somewhat lower performance will have other positive features which counteract the drawbacks, as described later. In the actual process, a high aspect ratio RIE process is first used to release the mechanical devices, and a following isotropic etch removes the underlying substrate material. A set of design rules assures that structures which are meant to be released are completely under-etched. A CMOS metal layer masks the areas where etching should be avoided. Any one of the five metal layers can be used as the top metal layer and will behave as a mask that determines the thickness of the resulting mechanical structure. The active CMOS circuit area has to be completely covered by metal. The designer must cope with two sets of geometrical design rules, e.g. one set for fulfilling the CMOS restrictions and another set for the MEMS part. More details can be found in [5].

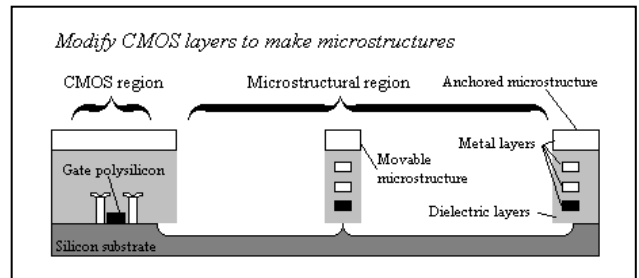

*Figure 1. Cross-section view of the post-CMOS process where the CMOS layers are used to make microstructures [6].*

### 3. A CMOS-MEMS OSCILLATOR

A CMOS-MEMS oscillator has been designed according to the ASIMPS procedure. The system consists of a vibrating cantilever beam coupled in loop with a feedback Pierce CMOS amplifier. Figure 2 shows the block diagram of the oscillator with its resonating MEMS beam and the feedback amplifier. The system will oscillate at a frequency given by the characteristic resonating mode of the cantilever beam modified by the input and output capacitances of the Pierce amplifier. If the beam is intentionally being bent by an external force, such as an acceleration, the spring coefficient of the beam is slightly altered which in turn changes the resonating frequency. Thus, acceleration can be measured by observing the frequency change.

The MEMS cantilever beam is designed to consist of a laminated structure of four aluminum metal layers separated by dielectric layers of $SiO_2$. A maximum thickness (lateral beam width) of 4.8 μm is obtained when using four metal layers in the ST7RF process. The width (or lateral height) of the beam is 2 μm and the length is chosen for a convenient resonance frequency and measurement resolution. Different versions of the oscillator have been designed.

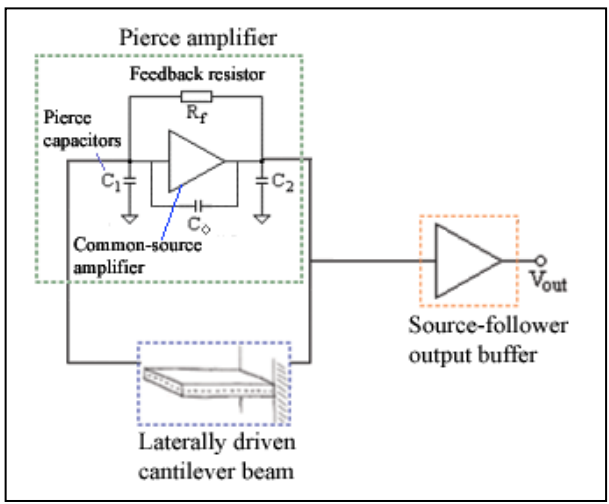

*Figure 2. Overview of the system containing the MEMS cantilever beam, a CMOS Pierce amplifier and an output buffer.*

An important parameter that decides the performance of the vibrating beam is the motional resistance $R_x$ [7]. The governing equation for the motional resistance of the cantilever beam is shown in equation 1

$$R_x = \frac{k}{\omega_0 Q \eta^2} \quad (1)$$

where k is the effective spring stiffness of the beam, $\omega_0$ is the resonance frequency, Q is the Q-factor and $\eta$ is the electromechanical coupling coefficient. A small gap between the electrode and the cantilever beam will increase the $\eta$ value. A low motional resistance is desirable, however the lateral width of the lateral beam is limited to 4.8 µm when using four metal layers as structural layers. As a result of a limited gap spacing and beam width, $R_x$ is a critical parameter for being able to initiate natural oscillation.

The CMOS part of the system consists of a Pierce amplifier and an accompanying bias network. The Barkhausen criterion [8] is used, stating that as long as this negative resistance is larger than $R_x$, natural oscillation of the system will occur. In order for oscillation to start up, the negative resistance $Re(Z_C)$ of the Pierce amplifier must be at least three times larger than the $R_x$ value. The negative resistance is controlled by tuning the Pierce capacitors $C_1$ and $C_2$, shown in figure 2. It is possible to adjust the parasitic $C_0$ capacitance by purposely routing the input and output of the amplifier close to each other. Equation 2 shows how to control the negative resistance [8]. $g_m$ is the transconductance of the common-source amplifier.

$$Re(Z_C) = \frac{g_m C_1 C_2}{(g_m C_0)^2 + \omega_0^2 (C_1 C_2 + C_2 C_0 + C_0 C_1)^2} \quad (2)$$

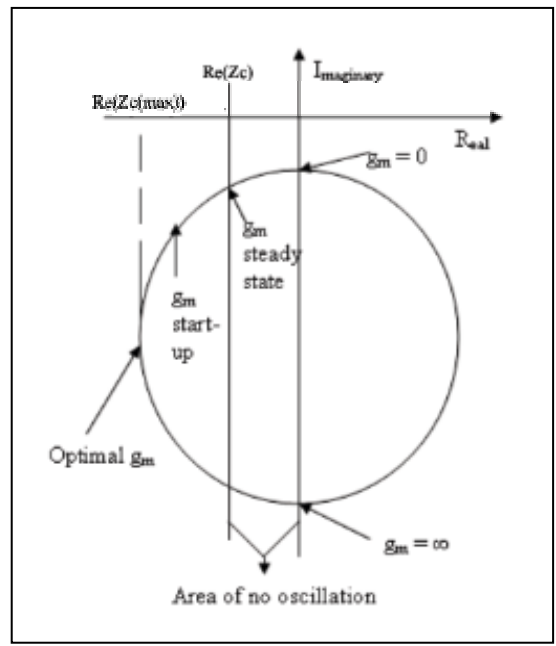

*Figure 3. Complex plane representation of the negative resistance.*

Figure 3 shows how equation 2 can be used to tune the location of the $Re(Z_C)$ compared to the maximum obtainable value $Re(Z_{C(MAX)})$. By using a tunable bias network, it is possible to tune the $g_m$ value in order to achieve the desired $Re(Z_C)$. The Pierce amplifier consists of a common-source (CS) amplifier with a tunable bias network as shown in figure 4. Transistors M1, M2 and M5 supply the bias current for the CS amplifier M7.

Transistor M6 in figure 4 acts as a high impedance feedback resistance. C1 and C2 are the accompanying Pierce capacitors which have been set to a value of 2pF in order to have low power consumption and little frequency pulling [9]. The Pierce oscillator is shown as the pink rectangle at the top left in figure 5. It can be seen that the MEMS beam takes up most of the space in addition to the passive components which deal with the decoupling of AC and DC. Those are required since this system is based on a one-port method. As the MEMS beam is to be used to detect external acceleration, the bending of the beam will be limited to 33% of total gap size. If the structure was to be used as a mixer oscillator instead, it is possible to make a two-port solution with the two electrodes on each side of the beam without the need for decoupling components. The negative effect of that solution is that it limits the bending of the beam to only 11% of the total gap size [10].

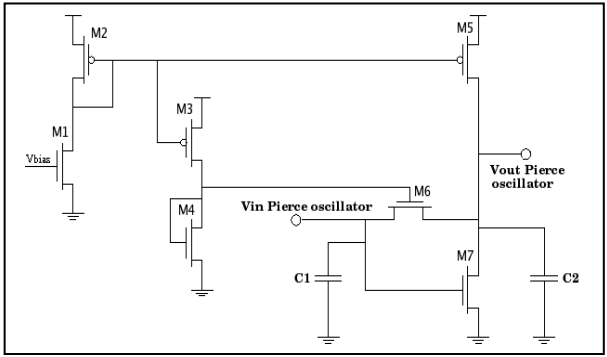

*Figure 4. Schematic view of common-source amplifier.*

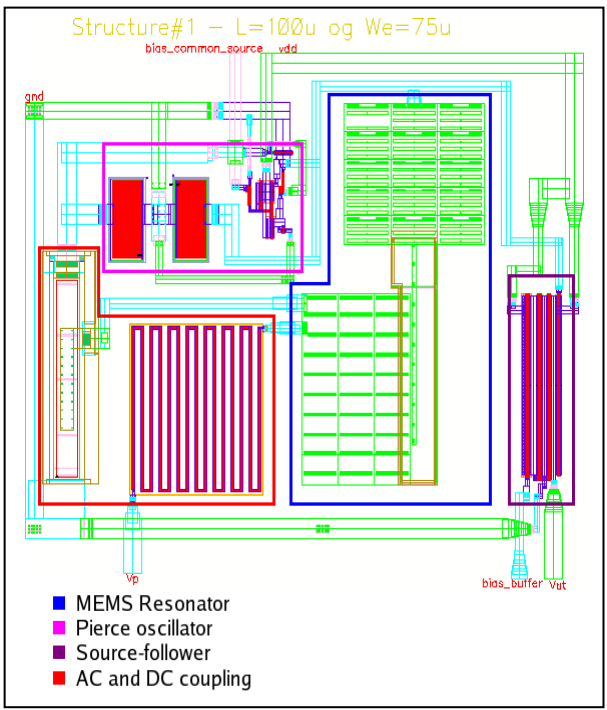

*Figure 5. Layout overview of design #1.*

Based on structural and dimensional choices, the electromechanical parameters for the MEMS part have been calculated. The lateral width of the beam is 4.8μm. The 1.2 μm gap between the drive electrode and the cantilever beam is critical for the resulting motional resistance $R_x$. The system operates with a $R_x$ of 717 kΩ. A total of three designs have been made using this post-CMOS technique. Relevant parameters for the three designs are shown in table 1.

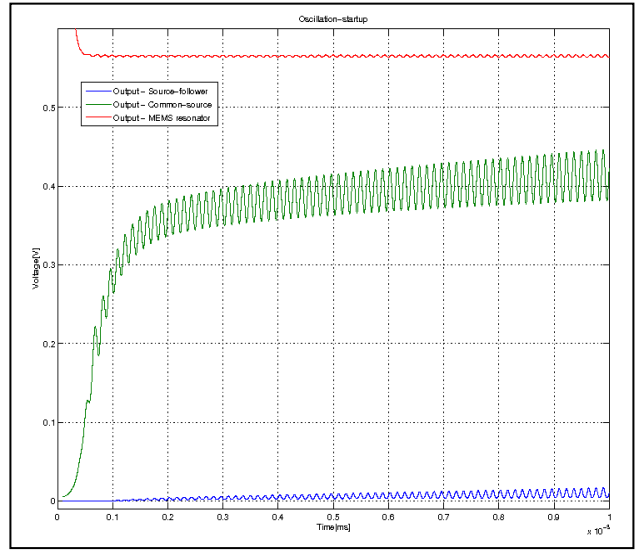

*Figure 6. Start-up simulation results using Cadence.*

|  | Design #1 | Design #2 | Design #3 (CC-beam) |
|---|---|---|---|
| $f_0$ [kHz] | 75.9 | 105.4 | 303.6 |
| $V_{pi}$ – pull-in [V] | 9.8 | 9.7 | 26.9 |
| $I_x$ - motional current[nA] | 3.4 | 2.9 | 1.5 |
| z–deflection [nm] | 161.1 | 171.2 | 22.0 |
| Re(Zc) [MΩ] | 64.7 | 33.6 | 4.7 |
| Re(Zc(max))[MΩ] | 103.8 | 74.7 | 25.9 |
| $R_x$ [kΩ] | 717.0 | 737.6 | 1008.3 |
| $L_x$ [H] | 6013.7 | 5011.5 | 2642.8 |
| $C_x$ [aF] | 731.1 | 454.8 | 104 |
| **Dimensions:** |  |  |  |
| H [μm] | 2 | 1 | 1 |
| W [μm] | 4.8 | 4.8 | 4.8 |
| L [μm] | 100 | 60 | 100 |
| $W_e$–electrode length [μm] | 75 | 45 | 80 |
| g – gap [μm] | 1.2 | 1.2 | 1.2 |

*Table 1. Relevant design parameters.*

Design #1 resulted in a 75.9 kHz oscillating frequency with a 3.4 nA current out from the MEMS resonator. Figure 6 shows how the circuit starts up naturally from thermal noise in the transistors. As Rx and $Re(Z_C)$ becomes equal, the amplitude of oscillation will stabilize.

# 4. FROM MEMS DEVICES TO SMART INTEGRATED SYSTEMS

## 4.1. General observations

The experiences from the design of the CMOS –MEMS oscillator have given us insight in possibilities, design constraints and problems of this procedure. Table 2 summarizes some of the general observations found by using the described CMOS-MEMS approach. Using an ordinary CMOS process certainly limits the possibility of what kind of mechanical structures could be implemented, e.g. which operation principles, materials and dimensions that can be used. Some parts of a composite system will most likely be implemented by non-optimal components with lower performance as compared to what could have been achieved by a customized MEMS process. However, even if the performance of single parts of the system will be lower, the overall system performance could still be of the same order ("good enough"). In our case integrated processing power can be used to compensate for the lack of performance for the single components and possible degradation of the MEMS components. The idea that an ensemble of non-optimal units can result in high performance by cooperative actions is analogous to what happened for integrated circuits some decades ago. There the integrated transistor, although having a poorer performance than the vacuum tube or its discrete variants, soon overtook the dominance and gave the possibility to develop real advanced information systems.

As can be generalized from the current example the MEMS structures typically move laterally and are specified geometrically. The procedure is not appropriate for vertical moving structures since it does not allow an electrode to be placed at a definite distance underneath. The isotropic etching will result in a large, not well defined distance between the metal layers and the underlying substrate. The resulting spacing is determined by the etching time, temperature and concentration of the etchant, all which is difficult to control to any absolute degree. Although somewhat restricted, this lateral operation principle is very flexible. If overhearing and cross-coupling of signals are problems, as for instance when implementing mixers, the mechanical structure can be organized to allow electrodes to be placed at some lateral distance. Reference [11] shows how electrode pairs for two different frequencies are conveniently separated.

The lateral gap dimension is very critical related to the performance and coupling efficiency between the mechanical and electrical domains. The minimum spacing that can be achieved is given by the resolution of the RIE etching. During that etching some polymer coating is added to the sidewalls, affecting the gap dimension. This means that the design rules have to take this extra coating into account. Some experiments at CMU, e.g. [12], showed that the gap between MEMS structures could be reduced after processing by including gap-closing structures in the design. By such an approach, an electrode can be slightly moved in lateral direction closer to another electrode by using thermal expansion.

| MEMS operation principles | Vertical movement:<br>- no well defined bottom electrode<br>Lateral movement:<br>+ movable structures defined by layout<br>+ flexible constructs<br>- lateral gaps given by MEMS design rules |
|---|---|
| MEMS structures | + geometry specified by layout<br>+ thickness defined by layout<br>- reduced stiffness<br>- stress mismatch<br>+ stress compensating design is possible |
| Design process | + design activity is well separated from implementation<br>+ layout oriented<br>+ offers an ASIC designer´s way of operating<br>+ library of standard cells can be used |
| MEMS layout rules | + securing complete release etching or anchoring<br>- resolution dependent of post CMOS RIE etching<br>- MEMS will take up costly CMOS space |
| CMOS – MEMS integration | + System-on-Chip is possible<br>+ short wires<br>+ low parasitics<br>+ low power consumption<br>+ easy interfacing |
| Implementation | CMOS process:<br>+ multiple choices, second sourcing<br>+ performance scales as processes develop<br>CMOS post processing:<br>+ low complexity, just a release step<br>+ no extra masking<br>+ can be performed on single chips<br>+ no complete wafers are needed |

*Table 2: General observations.*

The material properties (e.g. stiffness, Youngs modulus) of a laminated MEMS component are given by how many and which ones of the layers from the CMOS process are included in the layout. Omitting one or more of the layers will result in thinner structures. The designer will then have a specific set of possible thicknesses available. Since etching is avoided by covering areas by metal, the designer can choose the thickness of a MEMS beam by determining which of the metal layers should be the top shielding layer. For the current multilayer approach, a stress mismatch will occur due to different stress factors in the layers of the composite structures. The released structures then tend to curl, an effect which is more pronounced the thinner the structure. This curling might be critical and destroy the operation if a long thin beam departs from its electrodes. One way of reducing this adverse effect is by clever design to make both the electrodes and moving beams curl in the same direction and preferably to the same extent, by attaching the electrodes to a frame [13].

The design process that has been used, can be characterized as a layout process where the manipulation of geometrical entities is the central activity, - very much in the same way as has been the case in microelectronics for years. A set of MEMS layout rules had to be followed, and the performance could to a great extent be determined from the layout without coping with a lot of detailed processing parameters. Designing according to geometrical design rules and using standardized processes corresponds to what has been achieved for microelectronics, where the design activity is well separated from the processing.

One of the greatest advantages of this approach is the great flexibility the designer has for making optimal connections by utilizing diverse metal layers and vias. The interconnecting distance between the micromechanical and microelectronic components will be small, introducing very low parasitics compared to the impedance level to be handled when going off-chip. The post processing used in the described procedure does not need any extra masking. This is very important since an extra masking step would introduce post processing problems, the need for special "holders" etc. As the maskless etching is done on single chips, one is independent of post processing whole wafers.

## 4.2. The feasibility of the approach

Future smart systems are expected to contain devices for both sensing, signal adaptation, amplification and wireless RF transmission, - requiring a combination of different technologies. MEMS components expand the possibilities for the system designer, and a variety of both electronic and mechanical components should preferably be included in the same design process. Using the type of approach as described will leverage the way MST systems are designed and implemented in the future. The design process has been transferred to a geometrical one, manipulating geometries according to a specific set of rules. IPs and standard cells can be used to speed up the design process. The designer can take advantage of the large investments and developments in CMOS technology. Each new generation of the CMOS processes will offer new possibilities for increased performance. Combined with a simple release etch post processing this will allow a more standardized way of making combined MEMS and CMOS systems.

A special important and promising area for CMOS-MEMS systems would be to make compact nodes in wireless sensor networks. CMOS implementations of current RF nodes would need discrete off-chip components to obtain the required performance, e.g. inductors, varactors, oscillators and switches, [14]. Components made by RF MEMS technology are candidates for replacing those off-chip units. By the given method the RF MEMS parts can be effectively integrated with the microelectronic devices. Future transceiver nodes will be dependent on programmable features to cope with multi transmission standards and channels. Thus, electronics is needed for selecting the right MEMS units in each situation.

The described approach allows coils or inductors to be made as CMOS-MEMS. The resistance in a coil is minimized by coupling numerous metal layers together. The etching underneath effectively reduces the coupling to substrate, thus reducing eddy currents and capacitive loading. Likewise, capacitors for LC-tanks can be made by combs tuned by electro thermal actuation, [15]. True resonating structures of beams and lateral moving structures of different forms can be combined by CMOS amplifiers to implement oscillators, mixers and signal generators with reasonable Q-factors. The given example is a type of circuit which can be further optimized for such an application. Even if the Q-factors of pure poly silicon beams or disks can reach tens of thousands, lower values in CMOS-MEMS might suffice for given applications, especially because the combination with microelectronics is at hand. Specific signal processing and filtering can be done completely in the mechanical domain by connecting micromechanical components in proper ways [16]. This can be used to create exactly which filter functions are needed. Signal processing or mixing of frequencies can be done very efficiently in the mechanical domain before connecting to the electrical world. To select proper filter bands either filter banks made of fixed resonators selected by switches can be used, or tunable filters made of inductors and tunable capacitances, [17].

## 5. CONCLUSION

The paper has shown the feasibility of making combined MEMS and microelectronics using an ordinary CMOS process with its inherent design rule set. The simplicity and the low-cost post-CMOS processing method will to a great extent outweigh the disadvantages that exist. The project is a step towards designing future smart, robust, systems where the MST systems must be reprogrammed and reconfigured in flexible ways. This ability is central for making pervasive and ubiquitous computing systems which can be used for fine-grained observation and control. Only imagination will restrict the system modules which can be implemented and the applications made possible.